\documentclass[twocolumn,prd]{revtex4}
\usepackage{graphicx}

\begin{document}

\title{Cosmological Lower Bound on Dark Matter Masses 
from the Soft Gamma-ray Background}

\author{Kyungjin Ahn}
\author{Eiichiro Komatsu}
\affiliation{
Department of Astronomy, University of Texas at Austin\\
1 University Station, C1400, Austin, TX 78712
}

\date{\today}

\begin{abstract}
  Motivated by a recent detection of 511~keV photons from the center 
  of our Galaxy, we calculate the spectrum of the soft $\gamma$-ray background
  of the redshifted 511~keV photons from cosmological halos. 
  Annihilation of dark matter particles into electron-positron pairs
  makes a substantial contribution to the $\gamma$-ray background.
  Mass of such dark matter particles must be $\lesssim 100~\mathrm{MeV}$
  so that resulting electron-positron pairs are non-relativistic.
  On the other hand, we show that in order for the annihilation not to exceed 
  the observed background, the dark matter mass needs to be $\gtrsim 20~\mathrm{MeV}$. 
  We include the contribution from the active galactic nuclei and supernovae. 
  The halo substructures may increase the lower bound to $\gtrsim 60~\mathrm{MeV}$.
\end{abstract}


\maketitle

Recently, a narrow 511~keV $\gamma$-ray line from the central part of
our Galaxy
has been detected and mapped by the SPI spectrometer on the 
INTErnational Gamma-Ray Astrophysics Laboratory (INTEGRAL) satellite\cite{spi}.
This line should be produced by annihilation of non-relativistic
electron-positron pairs, and 
one of the possible origins is the
dark matter particles annihilating into 
electron-positron pairs\cite{mev_dm},
which explains the measured injection rate of positrons and 
morphology of the signal extended over the bulge region.
Therefore, 511~keV lines from our Galaxy as well as other galaxies 
may provide a smoking-gun for the existence of light dark matter 
particles\cite{hooper/etal:2004}.
Motivated by this idea, we calculate the spectrum of possible 
hard x-ray/soft $\gamma$-ray background (at $h\nu \sim 10-511$~keV) from the 
dark matter annihilation within cosmological halos distributed over a 
large redshift range.
\cite{substruct} and \cite{elsaesser} have considered similar 
calculations at much higher energy ($\gtrsim 1$~GeV), assuming that 
the signal is coming from annihilation of neutralinos whose mass
is typically of order $30-1000$~GeV; thus, their calculations cannot
be applied to the soft $\gamma$-ray background which 
we consider in this paper.
Also, we improve upon the previous calculations of this sort by explicitly
including the free-streaming and the Jean's mass scale of dark matter
particles, which 
will naturally (rather than arbitrarily) set the lower bound to the mass of 
dark matter halos.

The soft $\gamma$-ray background has been measured by
the A4 low energy detectors and
the A4 medium energy detectors on the HEAO-1 (High
Energy Astronomy Observatories 1) satellite as well as
a balloon experiment\cite{observation}.
Therefore, by comparing our calculations to observations,
one can constrain the properties of dark matter. 
As the line is narrow, the signal must come from 
non-relativistic electron-positron pairs; thus, 
the mass of dark matter particles 
must satisfy $511 \mathrm{keV}\lesssim m_X\lesssim 100$~MeV\cite{mev_dm}.
We calculate the background intensity, $I_{\nu}$, as \cite{peacock}
\begin{equation}
 I_{\nu} =
 \frac{c}{4\pi} 
 \int 
 \frac{dz\, P_{\nu}([1+z]\nu, z)}{H(z) (1+z)^{4}},
\label{eq-inu-generic}
\end{equation}
where $\nu$ is an observed frequency, $H(z)$ is the expansion rate at
redshift $z$, and 
$P_{\nu}(\nu, z)$ is the volume emissivity of 511~keV photons
in units of energy per unit time, per unit frequency and per unit
proper volume: 
\begin{equation}
  P_{\nu}=\delta\left((1+z)\nu-\nu_{511}\right)
  \alpha_{511}h\nu_{511} \langle \sigma v\rangle n^2_X.
\end{equation}
Note that an electron-positron pair creates two 511~keV photons,
and we assume that the universe is optically thin at $h\nu \gtrsim
10~\mathrm{keV}$.  
A parameter $\alpha_{511}$ determines the fraction producing 
an electron-positron pair per one dark matter annihilation process.
If annihilation occurs predominantly via positronium formation,
then $\alpha_{511}=1/4$. 
While it has been shown that the positronium formation is the dominant 
process at the center of our Galaxy\cite{kinzer/etal:2001}, we assume that
annihilation occurs directly, $\alpha_{511}=1$, for other galaxies.
Here, $h\nu_{511}=511$~keV and $\langle \sigma v\rangle$ is the thermally 
averaged annihilation cross section.
For $\Omega_{X}h^2=0.116$ \cite{wmap}
one finds $\left\langle \sigma v\right\rangle
= [3.4, 2.4, 2.9]\,10^{-26}~\mathrm{cm^{3}~s^{-1}}$ 
(or [1.1, 0.80, 0.97]~pb) for
$m_{X}=[1,\,10,\,100]~\mathrm{MeV}$, respectively
(e.g. Eq.~(1) in \cite{bes}). Here, we assume that 
$\langle \sigma v\rangle$ is velocity-indepedent (S-wave annihilation).
Boehm et al.\cite{mev_dm} argue that the S-wave cross section overpredicts 
the flux from the Galactic center; 
however, we argue that it is still consistent with the data for
$m_X\gtrsim 20$~MeV and $\rho\propto r^{-0.4}$ (or shallower). 
This profile explains morphology of the Galactic signal\cite{mev_dm}, and
one then finds that 
$\langle \sigma v\rangle \sim 0.8~{\rm pb}(m_X/20~{\rm MeV})^2$
is allowed at the 2-$\sigma$ level 
(see Eq.~(9) in \cite{hooper/etal:2004} where the r.h.s. 
should be multiplied by $37.6/17.3\simeq 2.2$ for $\rho\propto r^{-0.4}$).
As we show later, constraints from $\gamma$-ray background 
favor this parameter range, and our results do not violate the 
Galactic constraint.

Since the number density of dark matter particles, $n_X$, is usually
unknown, we use the mass density, $\rho_X\equiv n_X \, m_X$, instead of
$n_X$. After multiplying by $\nu=\nu_{511}/(1+z)$, one obtains
\begin{eqnarray}
\nu I_{\nu} 
 \nonumber
 &=& \frac{c}{H(z)(1+z)^4}\cdot\frac{h\nu_{511}\left\langle
 \sigma v\right\rangle }{4\pi m_{X}^{2}}
 \left\langle\rho_{X}^{2}\right\rangle_{z} \\
 &\simeq&  
\nonumber
0.4179~{\rm keV~cm^{-2}~s^{-1}~str^{-1}} \\
&\times&
\frac{C_X(z)(1+z)^2(\Omega_X h^2)^2}{\sqrt{\Omega_{m}h^2(1+z)^3
+\Omega_\Lambda h^2}}\\
\nonumber
&\times&
\left(\frac{\left\langle\sigma v\right\rangle}{10^{-26}~{\rm
 cm^{3}~s^{-1}}} \right)
\left(\frac{1~{\rm MeV}}{m_X}\right)^2,
\label{eq-nuinu}
\end{eqnarray}
where $1+z=\nu_{511}/\nu$, $\left\langle\rho_{X}^{2}\right\rangle_{z}$
is $\rho_{X}^{2}$ averaged over proper volume at $z$,
and $C_X(z)\equiv \left\langle\rho_{X}^{2}\right\rangle_{z}/\left\langle
\rho_{X}\right\rangle_{z}^{2}$ is the dark matter clumping factor.
(We have used $\left\langle\rho_{X}\right\rangle_z=10.54~\Omega_X
h^2(1+z)^3~{\rm keV~cm^{-3}}$.) 
It is worth noting that since $\left\langle\sigma v\right\rangle\propto 
(\Omega_X h^2)^{-1}$\cite{kolbturner}, the intensity scales
approximately as $ \nu I_\nu\propto \left[C_X(z)/m_X^2\right] \left( \Omega_X
 /\Omega_{m}^{1/2} \right) h(1+z)^{1/2}$
for $z > 1$; thus, the larger the mass per dark 
matter particle is, the smaller the predicted $\gamma$-ray background
becomes ($\propto m_X^{-2}$),  
and cosmology and redshift (or $\nu$) dependence of the signal
is almost entirely determined by $C_X(z)$, i.e., how clumpiness of the
dark matter 
halos evolves with redshift.
Throughout this paper, we adopt the best-fit power-law $\Lambda$CDM
model from the first year Wilkinson Microwave Anisotropy Probe ({\sl
WMAP}) satellite \cite{wmap}. 

Eq.~(\ref{eq-nuinu}) allows one to calculate
the intensity directly from $N$-body simulations which, in principle, give
the full evolution of $C_X(z)$; however, such simulations are computationally
challenging as they must resolve individual halo profiles at small distances 
(e.g., $\sim {\rm kpc}$) in a large simulation box (e.g., $\sim 1~{\rm Gpc}$), 
which requires a large number of particles. 
Instead, we take an alternative approach. The clumping factor is
essentially determined by  
the clumping of individual halos and the volume occupied by halos (or
the number of halos).  
The former can be obtained by simulating individual halos using
high-resolution simulations  
with smaller box sizes, while the latter can be obtained by simulating
many halos using  
low-resolution simulations with larger box sizes.
Therefore, by combining these simulations, one can calculate the
clumping factor. This somewhat  
empirical approach, called a halo approach (see \cite{cooray_sheth}
for a review), 
turns out to be very powerful in making accurate predictions for clustering
of dark matter particles in a highly non-linear regime. By using this
approach, we calculate  
$C_X(z)$ as
\begin{equation}
  C_X(z) =
 \frac{(1+z)^{3}}{\langle\rho_X\rangle_z^2} \int_{M_{min}}^{\infty}
 dM \frac{dn(M,z)}{dM}
 \int d^{3}r\rho_{X}^{2}(M,r),
 \label{eq-rhox2}
\end{equation}
where $\rho_X(M,r)$ is a dark matter halo profile, 
and $dn(M,z)/dM$ is the comoving number density of halos in the mass range of
$M\sim M+dM$ at $z$. For $dn/dM$ we adopt an empirical fit to $N$-body
simulations obtained  
by \cite{sheth} with $a=0.75$, $p=0.3$, and $A=0.322$ (see
\cite{sheth} for the detail). 
Why does the integral in Eq.~(\ref{eq-rhox2}) have a minimum mass?
When the length scale is smaller than the free-streaming scale and/or 
Jean's scale of the dark matter particles, density fluctuations
are suppressed and halos do not form. Although the actual effect is
not an abrupt cut-off  
and will cause a gradual suppression of halos toward low mass scales,
we approximate 
it as the low mass cut-off in the integral; thus, no halo would form below 
$M_{\rm min}=max(M_{F},\, M_{J})$, where $M_{F}$ is the free-streaming
mass and $M_{J}$ is  
the dark matter Jean's mass. For $M_{F}$, we adopt the expression
obtained by \cite{mf}: 
$M_{F}\approx(4\pi/3)\overline{\rho_{m}}(\lambda_{F}/2)^{3}$,
where $\lambda_{F}\equiv
2.72(\Omega_{X}h^{2})^{1/3}(m_{X}/1\mathrm{keV})^{-4/3}$~Mpc, and 
$\overline{\rho_{m}}$ is the present-day mean matter density.
With this definition one obtains \cite{mf}
\begin{equation}
  M_{F}=8.3\cdot10^{-2}~M_{\odot}\left(\frac{\Omega_{m}h^{2}}{0.13}\right)
  \left(\frac{\Omega_{X}h^{2}}{0.11}\right)\left(\frac{m_{X}}
  {\mathrm{1MeV}}\right)^{-4}.
\label{eq-mf-num}
\end{equation}
The dark matter Jean's mass is given by \cite{cl}
\begin{equation}
  M_{J}=38.79~M_{\odot}\left(\frac{m_{X}}{\mathrm{MeV}}\right)^{-3}
  \left(\frac{x_{F}}{12}\right)^{3/2}\left(\frac{1+z}{3069}\right)^{3/2},
\label{eq-mj-num}
\end{equation}
for $1+z>1+z_{eq}\simeq 3069$, which is a valid expression for cold thermal
relics which decouple from radiation at 
$x_{F}=m_{X}/T_{F}>3$
($x_{F}\approx[11-15]$ for $m_{X}\approx[1-100]~\mathrm{MeV}$;
e.g. see \cite{bes}).

\begin{figure}
\includegraphics[width=86mm]{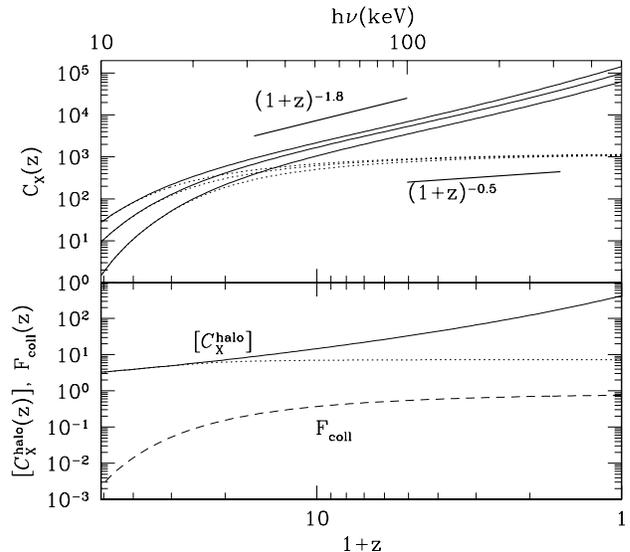}
\caption{\label{fig-clump}
  {\it Upper panel}: The clumping factor $C_{X}(z)$ for different $m_{X}$'s. 
  Solid lines, from
  the bottom to the top, represent the clumping factor for
  $m_{X}=[1,10,100]\mathrm{MeV}$ respectively. Dotted lines represent
  the clumping 
  factor for $c_{\rm max}=3$, also with the same meaning from the bottom
  to the top. Two power laws reflecting the asymptotic behavior of
  $C_{X}(z)$ are also plotted. The trend for different mass
  curves indicates the fact that the minimum halo mass is smaller for
  larger $m_{X}$, thus obtaining higher clumping. 
  {\it Lower panel}: The mean {\it halo} clumping factor $[{\cal
  C}_{X}^{\rm halo}(z)]$
  and  the collapse fraction $F_{{\rm coll}}(z)$ for $m_{X}=1~{\rm MeV}$. The
  solid line is for no upper limit on $c$, while the dotted line is
  for $c_{\rm max}=3$. It
  clearly shows that the low redshift behavior is determined mainly by
  $[{\cal C}_{X}^{\rm halo}(z)]$, while the high redshift behavior by $F_{{\rm
  coll}}(z)$. 
}
\end{figure}

Eq.~(\ref{eq-rhox2}) can be cast into the product of the collapse
fraction and the mean ``halo clumping factor'':
\begin{equation}
  C_X(z) =
  \Delta \cdot F_{{\rm coll}}(z) \cdot \left[ {\cal C}_{X}^{\rm halo}
  \right],
\label{eq-cx}
\end{equation}
where $\Delta$ is the mean halo
overdensity of a halo in units of the cosmic mean matter density,
the 
collapse fraction $F_{{\rm coll}}(z) \equiv \int dM 
\frac{dn}{dM} M /  \overline{\rho_{m}} $ is
the mass fraction 
collapsed into cosmological halos at $z$, 
$[A]\equiv \int dM \frac{dn}{dM}MA / \int dM \frac{dn}{dM}M$, and
${\cal C}_{X}^{\rm halo}\equiv \int d^{3}r \left(
\frac{\rho_{X}}{\left\langle \rho_{X} 
\right\rangle_{halo}} \right)^{2} / \int d^{3}r$ is the ``halo
clumping factor'' defined in terms of the halo mean density 
$\left\langle \rho_{X}\right\rangle_{halo}\equiv \int d^{3}r \rho_{X}
/ \int d^3{r} = \Delta \cdot \left\langle \rho_{X}\right\rangle_{z}$. As
illustrated in Fig.~(\ref{fig-clump}), the early 
time $C_{X}(z)$ is mainly determined by $F_{\rm coll}(z)$, while the
late time $C_{X}(z)$ by $[{\cal C}_{X}^{\rm halo}]$.
As it can be seen from Eq.~(\ref{eq-rhox2}), $C_X(z)$ is very
sensitive to the density profile, whose properties are not fully
understood (or observed) yet.
In order to quantify uncertainties associated with the density profile,
therefore, we adopt two different models for density profiles.

{\it Case A: The Navarro-Frenk-White (NFW) Profile.}
The NFW profile is an empirical fit to radial profiles of dark matter halos 
in $N$-body simulations \cite{nfw}.
This profile has a central cusp, $\rho_X\propto r^{-1}$, and is therefore
expected to produce large annihilation signals.
The NFW profile is given by $\rho_X(r) = \rho_{s}(r/r_{s})^{-1}
(1+r/r_{s})^{-2}$\cite{nfw}. 
The mass $M$ enclosed within the virial radius, $r_{\rm vir}$, is
$M= 4\pi\rho_{s}r_{s}^{3}\left[\ln(1+c)-{c}/(1+c)\right]$,
and the scale radius, $r_{s}$, is $r_s=r_{\rm vir}/c$, where
$c$ is called the concentration parameter. With these definitions, one
obtains
\begin{equation}
{\cal C}_{X}^{\rm halo}=\frac{c^{3}(1-1/(1+c)^3)}{9(\ln(1+c)-c/(1+c))^2}.
\end{equation}
For instance, ${\cal C}_{X}^{\rm halo}(c=3)=7.3$,
${\cal C}_{X}^{\rm halo}(c=10)=50$, ${\cal C}_{X}^{\rm halo}(c=20)=203$.
For $\Delta$, we use an approximate form $\Delta=(18\pi^{2}+82 x -39 x^{2})/\Omega (z),$
where $\Omega(z)$ is the ratio of mean matter density to the critical
density at $z$, and $x=\Omega(z)-1$. This expression is valid for
a flat $\Lambda$CDM universe \cite{bryannorman,bullock}.
The concentration of dark matter halos found in $N$-body simulations 
has a log-normal distribution with a median value of \cite{bullock}
\begin{equation}
  c(M,z)=4\frac{1+z_{c}}{1+z},
\label{eq-c}
\end{equation}
where the collapse redshift, $z_c$, is implicitly given by a relation
$M_{*}(z_{c})=10^{-2}M$. ($M_*(z)$ is the non-linear mass at $z$.)
As lower mass objects collapse at higher $z_c$, the concentration decreases
as $M$. (The lower mass objects have steeper profiles.)
Following  \cite{substruct1}, we take into account
a log-normal distribution of $c$ with the dispersion of $\sigma(\ln(c))=0.2$.
While we use Eq.~(\ref{eq-c}) for all the range of $M$ and $z$ in our analysis,
we should keep in mind that this fitting formula is valid only for a limited 
range of $M$ and $z$ covered by $N$-body simulations.
\begin{figure}
\includegraphics[width=86mm]{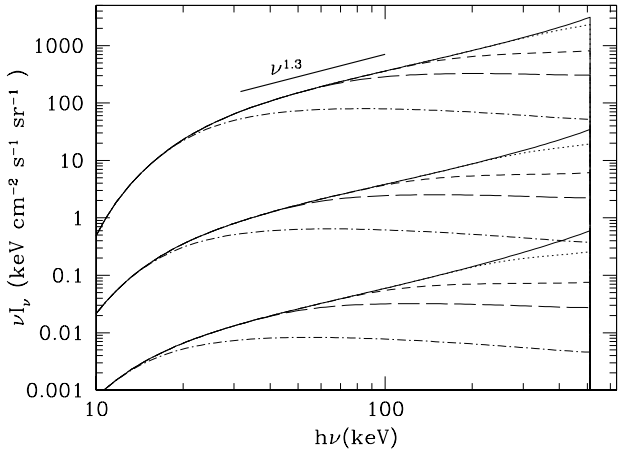}
\caption{\label{fig-mchi}
  Predicted spectrum of the soft $\gamma$-ray background from 
  cosmological halos with the NFW profile.
  From the top to the bottom, each set of curves represent 
  $m_{X}=1, 10, 100~\mathrm{MeV}$.
  The solid lines show the canonical concentration model
  [Eq.~(\ref{eq-c})], while 
  the dotted, short-dashed, long-dashed and dot-dashed curves show 
  the models with upper cut-offs of $c_{\rm max}=$50, 20, 10 and 3,
  respectively. Signals without upper cut-offs are well fit by a
  power-law for $\nu \gtrsim 40 \mathrm{keV}$, as indicated by the
  uppermost line.}
\end{figure}

\begin{figure}
\includegraphics[width=86mm]{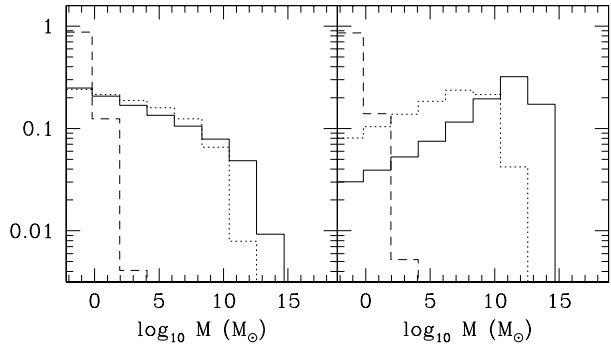}
\caption{\label{fig-histo}Fractional contribution
from different NFW halo masses at different redshifts. 
The left panel shows the 
canonical concentration model [Eq.~(\ref{eq-c})] with no upper limit, 
while the right panel shows the model with $c_{\rm max}=3$.
In both panels, the solid, dotted and dashed lines correspond
to $z=$0, 4 and 50 respectively.}
\end{figure}

Fig.~\ref{fig-mchi} shows the predicted soft $\gamma$-ray background
for the NFW profile. The predicted signal, $\nu I_\nu$, is roughly
proportional  
to $\nu^{1.3}$, which simply
reflects the fact that the universe becomes more clumpy at lower $z$
as $C(z)\propto (1+z)^{-1.8}$
(recall that $h\nu=511~{\rm keV}/(1+z)$), as seen in Fig.~\ref{fig-clump}.
One should, however, keep in mind that we have simply extrapolated 
the fitting formula for $c(M,z)$ [Eq.~(\ref{eq-c})] to the regime
where simulations 
are no longer valid. For the $m_{X}=1~\mathrm{MeV}$ case, for example,
the free-streaming  
mass gives the maximum concentration of $c\sim 70$ at $z=0$.
Since we don't fully understand halo profiles at such low mass or high
concentration, 
we also apply an arbitrary, hypothetical upper limit to $c$ 
and investigate sensitivity of our results to the change in
concentration parameters. 
This toy model will substantially reduce the contribution from small mass halos
with $c(M,z)>c_{\rm max}$. As $c(M,z)\propto (1+z)^{-1}$, the signal
at lower $z$  
(higher $h\nu$) is suppressed. As a result, $\nu I_\nu$ becomes almost flat
for $\nu>\nu_*$, where $\nu_*$ corresponds to a redshift, $z_*$, after which
the clumping factor stops evolving fast and evolves only slowly as 
$(1+z)^{-1/2}$ or even slower (Fig. \ref{fig-mchi}, \ref{fig-clump}).
Sensitivity of the predicted spectrum to the concentration parameter model 
is demonstrated more in Fig.~\ref{fig-histo}. For the canonical 
concentration parameter model given by Eq~(\ref{eq-c}) without any
upper limit,  
the smallest mass halos always dominate; on the other hand, once the upper 
limit on $c$ is imposed, the largest contribution comes from $M\simeq
M_{*}(z)$,  
effectively removing contribution from the lower mass halos as the
lower mass halos 
hit $c_{\rm max}$ earlier than the higher mass halos.

{\it Case B: The Truncated Isothermal Sphere (TIS) Profile.}
The TIS \cite{tis} is an analytical model of the post collapse
equilibrium structure of halos resulting from the collapse of a top-hat 
density perturbation. The TIS is the minimum energy solution to the 
Lane-Emden equation, matching energy of the TIS to the energy of the 
top-hat perturbation. The TIS density profile has a soft core and thus 
provides a much smaller annihilation signal than the NFW profile. Note
that the TIS fits  
the observed rotation curves of dwarf spheroidals and low surface brightness
galaxies fairly well, as opposed to the NFW. The TIS density profile is 
given by \cite{tis}
\begin{equation}
  \rho=1799\overline{\rho_{m}}(z)
  \left[\frac{21.38}{9.08^{2}+\left(r/r_{0}\right)^{2}}
  -\frac{19.81}{14.62^{2}+\left(r/r_{0}\right)^{2}}\right],
\label{eq-rho-tis}
\end{equation}
for $r<r_{t}$. Here, $\overline{\rho_{m}}(z)$ is the mean matter
density at $z$, $r_{t}$ is the truncation radius, 
\begin{equation}
  r_{t}=\frac{0.187}{1+z}\left(\frac{M}{10^{12}M_{\odot}}\right)
  (\Omega_m h^2)^{-1/3}~\mathrm{Mpc},
\label{eq-rt-tis}
\end{equation}
and $r_{0}$ is the core radius given by $r_{0}=r_{t}/29.4$.
We find that the predicted signal for the TIS model 
is very similar to that for the NFW with $c_{\rm max}=3$,
which is consistent with the TIS profile being much flatter (i.e.,
much less clumpy) than the NFW with the canonical concentration. This
can be more easily explained by Eq.~\ref{eq-cx}. At $z\gtrsim1$, 
$\Delta\approx 178$ and $C_{X}(z) \approx 178 \times 7.3 \times F_{\rm
  coll}$ for the NFW with $c_{\rm max}=3$, while the TIS has $\Delta
\approx 130$, 
${\cal C}^{\rm halo} \approx 11.5$, and so $C_{X}(z) \approx 130
\times 11.5 \times F_{\rm coll}$. 

\begin{figure}
\includegraphics[width=86mm]{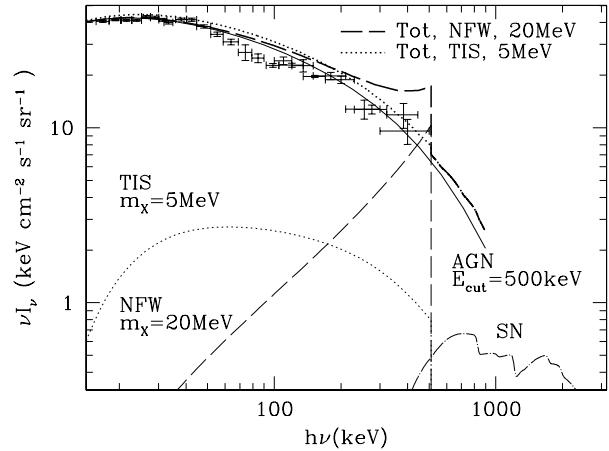}
\caption{\label{fig-total}
   Comparison of the dark matter annihilation (dashed lines for the NFW 
   and dotted lines for the TIS), the Compton-thin AGNs\cite{ueda}
   (solid), and the  
   type Ia supernovae contribution\cite{paperii} (dot-dashed), as well
   as the observed  
   soft $\gamma$-ray background\cite{observation} (points with errors). 
   The predicted total signals are plotted as the thick dashed and 
   dotted lines, which set upper limits on the dark matter mass
   from the measured points.}
\end{figure}

{\it AGNs and Supernovae.} There are other sources for the soft
$\gamma$-ray background. 
The Active Galactic Nuclei (AGNs) are probably the most dominant contribution 
up to $\sim 100$~keV, and it may continue to dominate up to
$\sim 500$~keV\cite{comastri,ueda}, depending on a cut-off
energy scale of the AGN spectrum, $E_{\rm cut}$, which is a free parameter.
The other candidate source is the Type Ia Supernovae (SNIa).
\cite{the} have calculated a spectrum of the soft $\gamma$-ray
background from SNIa, finding that the SNIa contribution can account
for the observed intensity. Our recent calculations, however,
show that they overestimated the signal by a factor of $\sim 10$
(they assumed $\sim 10$ times larger supernova rate than observed).
The SNIa contribution is thus negligible at the energy scale of 
our interest\cite{paperii}.
Figure~\ref{fig-total} compares the dark matter annihilation, 
the Compton-thin AGNs with $E_{\rm cut}=500~{\rm keV}$\cite{ueda}, 
and the supernovae contribution\cite{paperii}, as well as 
the observed soft $\gamma$-ray background\cite{observation}.
We find that an acceptable range for the dark matter mass 
is $20\lesssim m_{X} \lesssim 100~\mathrm{MeV}$ for the NFW profile.
(Note that the upper limit comes from requiring that electron-positron
pairs be non-relativistic.)
The uncertainty includes statistical uncertainty as well as potential systematic
uncertainties associated with instrumental calibrations. It has been shown
that the background signals measured by various experiments show a large
scatter (e.g., \cite{rxte}), and it may be possible that the measured background
is uncertain up to 30\% \cite{ueda}.

\begin{figure}
\includegraphics[width=86mm]{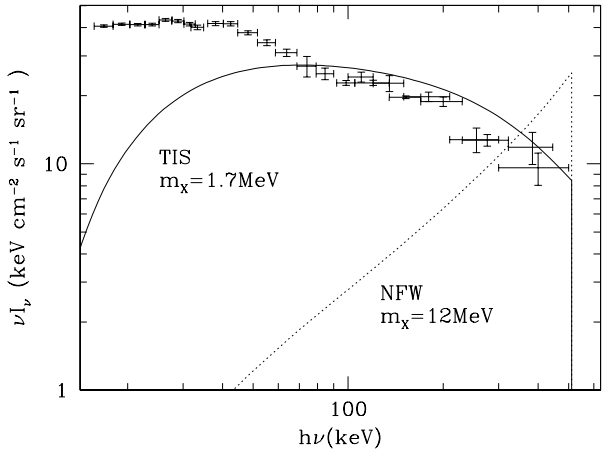}
\caption{\label{fig-dmonly}
  Dark matter annihilation only. 
  Smaller dark matter masses are allowed. (see also Fig.~\ref{fig-total}.)} 
\end{figure}

One may relax this constraint in various ways. If the AGN spectrum
has a smaller energy cut-off, $E_{\rm cut}\sim 100~{\rm keV}$, 
the AGNs do not contribute to the $\gamma$-ray background
for $h\nu \gtrsim 100~\mathrm{keV}$, and 
$m_{X} \gtrsim 12\,\mathrm{MeV}$ becomes 
acceptable (Fig.~\ref{fig-dmonly}). 
Bigger uncertainties come from the halo model. 
If we adopt the TIS or the NFW with $c_{\rm max}=3$,  
we obtain $m_{X} \gtrsim 5$ and 1.7~MeV for with and without the AGNs 
contribution, respectively (see Fig.~\ref{fig-total} and \ref{fig-dmonly}).
Such a low concentration parameter or filtering of low mass halos may
be possible if dark matter has some finite self-interacting cross
section\cite{spergel_steinhardt, ahnSIDM}. 
In addition to the average shape of $\rho_X$, the clumpy substructure
within halos would contribute to the signal
significantly\cite{substruct,substruct1}. 
While the abundance and properties of substructure are uncertain,
the analysis in \cite{substruct} shows that the substructures can
enhance the clumping factor 
by more than a factor of 10. 
As the intensity is proportional to $m_X^{-2}$, the contribution from
the substructures can increase the lower limit on $m_X$ by more than a factor
of 3, giving $m_X\gtrsim 60~{\rm MeV}$ for the NFW profile.

In summary, we have calculated the soft $\gamma$-ray background of the 
redshifted 511~keV photons from the dark matter particles annihilating 
into electron-positron pairs in cosmological halos.
Our fiducial model based on the current $N$-body simulations, 
compared to the observed $\gamma$-ray
spectrum, places limits on the dark matter mass, 
$20\lesssim m_{X}\lesssim 100~\mathrm{MeV}$. 
(If the cross section is allowed to be free, we find
$(\langle \sigma v\rangle/{\rm pb})({\rm MeV}/m_X)^2 
\lesssim 2.1\times 10^{-3}$, which is consistent with that from 
the Galactic center\cite{hooper/etal:2004}
for $\rho\propto r^{-0.4}$ or shallower profile).
Recently, Beacom, Bell and Bertone\cite{BBB} have shown that the internal
bremsstrahlung of electrons and positrons contributes 
to the $\gamma$-ray background at $>1~{\rm MeV}$.
Using COMPTEL and EGRET data, they have obtained $m_X\lesssim 20$~MeV.
Putting all the constraints together, $m_X\sim 20$~MeV may be favored.
However, in order to have a more robust
constraint, further understanding of cosmological nonlinear structures
is required. Obtaining a more reliable model for AGNs is equally important.
Uncertainty in the cut-off energy, $E_{cut}$, is the dominant uncertainty 
in the predicted soft $\gamma$-ray background signal at $h\nu 
\gtrsim 100~\mathrm{keV}$. 
If the cut-off energy is smaller than $\sim 100~{\rm keV}$, 
the soft $\gamma$-ray background signal may serve as evidence for
the existence of dark matter particles with $m_X\sim {\cal O}(10)$~MeV. 

We would like to thank D.E. Gruber for providing us with the HEAO-1 
data, 
and Y. Ueda for providing us with the AGN predictions.
We would like to thank E.R. Fernandez for bringing our attention to 
this subject, and G. Bertone, D.E. Gruber, R.E. Rothschild and P.R. 
Shapiro for valuable discussion.
K. A. was partially supported by NASA Astrophysical Theory Program 
grants NAG5-10825, NAG5-10826, NNG04G177G, and Texas Advanced Research 
Program grant 3658-0624-1999.



\begin{thebibliography}{99}
\bibitem{spi}P. Jean et al., Astron. Astrophys. {\bf 407}, L55
  (2003); J. Kn\"{o}dlseder et al., Astron. Astrophys. {\bf 411}, L457
  (2003) 
\bibitem{mev_dm}C. Boehm, D. Hooper, J. Silk, M. Casse and J. Paul,
  Phys. Rev. Lett. {\bf 92}, 101301 (2004)
\bibitem{hooper/etal:2004} D. Hooper, F. Ferrer, C. Boehm, J. Silk, J. Paul, N.W. Evans,
and M. Casse, Phys. Rev. Lett., {\bf 93}, 161302 (2004)
\bibitem{substruct}J. Taylor and J. Silk,
  Mon. Not. R. Astron. Soc. {\bf 339}, 505 (2003)
\bibitem{elsaesser}D. Els\"{a}esser and K. Mannheim, Astropart. Phys. {\bf
  22}, 65 (2004) 
\bibitem{observation}D.E. Gruber, J.L. Matteson, L.E. Peterson and
  G.V. Jung, Astrophys. J. {\bf 520}, 124 (1999); 
Y. Fukada, S. Hayakawa, I. Kasahara, F. Makino, I. Suzuki, Y. Tanaka and
  B.V. Sreekantan, Nature {\bf 254}, 398 (1975)
\bibitem{peacock} J.A. Peacock, \textit{Cosmological Physics} (Cambridge
  University Press, 1999), pp 91 - 94
\bibitem{kinzer/etal:2001} R.L. Kinzer, P.A. Milne, J.D. Kurfess,
M.S. Strickman, W.N. Johnson and W.R. Purcell, Astrophys. J. {\bf 559}, 282 (2001)
\bibitem{wmap}D.N. Spergel et al., Astrophys. J. Supp. Ser. {\bf 148}, 213
  (2003)
\bibitem{bes}C. Boehm, T.A. Ensslin and J. Silk, J. Phys. G {\bf 30},
  279 (2004) 
\bibitem{kolbturner}E.W. Kolb and M.S. Turner, \textit{The Early
  Universe} (Addison-Wesley, 1990), Eq (5.47)  
\bibitem{cooray_sheth} A. Cooray and R. Sheth, Phys. Rept. {\bf 372}, 1 (2002)
\bibitem{sheth} R.K. Sheth and G. Tormen,
  Mon. Not. R. Astron. Soc. {\bf 329}, 61 (2002)
\bibitem{mf}J. Sommer-Larsen and A. Dolgov, Astrophys. J. {\bf 551},
  608 (2001); V. Avila-Reese, P. Col\'{i}n, O. Valenzuela, E. D'Onghia and
  C. Firmani, Astrophys. J. {\bf 559}, 516 (2001)
\bibitem{cl}P. Coles and F. Lucchin, \textit{Cosmology} (John Wiley \&
  Sons, 1995), Eq (13.5.10)
\bibitem{nfw}J.F. Navarro, C.S. Frenk and S.D. White,
  Astrophys. J. {\bf 490}, 493 (1997)
\bibitem{bryannorman}G.L. Bryan and M.L. Norman, Astrophys. J. {\bf
  495}, 80 (1998)
\bibitem{bullock}J.S. Bullock, T.S. Kolatt, Y. Sigad, R.S. Somerville,
  A.V. Kravtsov, A.A. Klypin, J.R. Primack and A. Dekel,
  Mon. Not. R. Astron. Soc. {\bf 321}, 559 (2001) 
\bibitem{substruct1}P. Ullio, L. Bergstr\"{o}m, J. Edsj\"{o} and C. Lacey,
Phys. Rev. D {\bf 66}, 123502 (2002)
\bibitem{tis}P.R. Shapiro, I.T. Iliev and A.C. Raga,
  Mon. Not. R. Astron. Soc. {\bf 307}, 203 (1999); I.T. Iliev and
  P.R. Shapiro, Mon. Not. R. Astron. Soc. {\bf 325}, 468 (2001)
\bibitem{comastri}A. Comastri, G. Setti, G. Zamorani and G. Hasinger,
Astron. Astrophys. {\bf 296}, 1 (1995)
\bibitem{ueda}Y. Ueda, M. Akiyama, K. Ohta and T. Miyaji,
Astrophys. J. {\bf 598}, 886 (2003)
\bibitem{the} L.-S. The, M.D. Leising and D.D. Clayton,
Astrophys. J. {\bf 403}, 32 (1993)
\bibitem{paperii}K. Ahn et al., in preparation
(2004)
\bibitem{rxte} M. Revnivtsev, M. Gilfanov, R. Sunyaev, K. Jahoda and C. Markwardt,
Astron. \& Astrophys. {\bf 411}, 329 (2003)
\bibitem{spergel_steinhardt}D.N. Spergel and P.J. Steinhardt,
Phys. Rev. Lett. {\bf 84}, 3760 (2000) 
\bibitem{ahnSIDM}K. Ahn and P.R. Shapiro,
Mon. Not. R. Astron. Soc., submitted [arXiv:astro-ph/0412169]
\bibitem{BBB} J.F. Beacom, N.F. Bell and G. Bertone, Phys. Rev. Lett., submitted 
[arXiv:astro-ph/0409403]

\end{thebibliography}
\end{document}